\documentclass[aps,prl,notitlepage,twocolumn,superscriptaddress,longbibliography,nofootinbib]{revtex4-1}
\pdfoutput=1
\usepackage[utf8]{inputenc}

\usepackage{amsmath,amsthm,amssymb,amsfonts}
\usepackage{scalerel}
\usepackage{mathtools}
\usepackage{graphicx}
\usepackage{subfigure}
\usepackage[normalem]{ulem}
\usepackage[colorlinks = true,
            linkcolor = red,
            urlcolor  = magenta,
            citecolor = red,
            anchorcolor = blue]{hyperref}
\usepackage{mathrsfs}
\usepackage{bbold}
\usepackage{units}
\allowdisplaybreaks
\usepackage{bm}
\usepackage{braket}
\usepackage{makecell}
\usepackage[nameinlink]{cleveref} 
\usepackage{upgreek}
\usepackage{blindtext}
\usepackage{verbatim}
\usepackage{algorithm}
\usepackage[noend]{algpseudocode}
\usepackage[dvipsnames]{xcolor}
\usepackage{bbm}
\usepackage{array}
\usepackage{multirow}
\usepackage{tabularx}
\usepackage{float}
\usepackage{dcolumn}
\usepackage{slashed}
\usepackage{braket}
\usepackage{verbatim}
\usepackage{multirow}
\usepackage{resizegather}
\usepackage{titlesec}
\usepackage[toc,page]{appendix}
\usepackage[export]{adjustbox}

\newcommand{\tr}{\mathrm{tr}}
\theoremstyle{definition}

\theoremstyle{plain}
\newtheorem{prop}{Property}
\theoremstyle{plain}
\newtheorem*{prop*}{Property}

\newtheorem*{ctr*}{Conjecture}

\begin{document}

\title{Exact Entanglement Dynamics Beyond Nearest-Neighbor Dual-Unitary Floquet Systems \\
}
\author{Tanay Pathak\,\,\href{https://orcid.org/0000-0003-0419-2583}
{\includegraphics[scale=0.05]{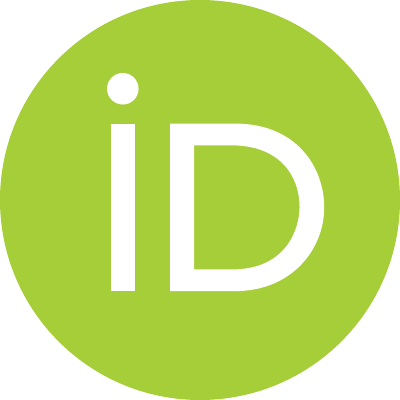}}}
\email{pathak.tanay.4s@kyoto-u.ac.jp}
\affiliation{Department of Physics, Kyoto University, Kitashirakawa Oiwakecho, Sakyo-ku, Kyoto 606-8502, Japan}

\begin{abstract}
Exact results using dual-unitarity largely rely on nearest-neighbor structures, while finite-range interactions typically lead to complications. Going beyond the usual nearest-neighbor setting, we introduce an analytically tractable family of finite-range kicked Ising models that admit exact closed-form entanglement dynamics. The construction is based on a staggered
structure in which dual-unitarity is present on sublattices that are then coupled to each other. The central observation is that these inter-sublattice couplings do not obstruct the dual-unitarity of the resulting model. For the minimal interaction range of $r= 2$, we derive exact expressions for all the $n-$R\'enyi entanglement entropies at all times and show that the result is the sum of the two coupled sublattice contributions. Our framework extends naturally to larger finite interaction ranges and to systems with heterogeneous local Hilbert spaces, without additional assumptions. It thus provides a controlled setting for studying exact entanglement growth beyond strictly nearest-neighbor dual-unitary models.
\end{abstract}

\maketitle


\textit{Introduction.} Quantum many-body phenomena are both fascinating and difficult to analyze, largely because the size of the Hilbert space grows exponentially with system size. As a result, understanding the time evolution of observables is challenging, even with numerical methods. One useful approach to this problem is based on space-time duality, which exploits a discrete duality between spatial and temporal directions to study quantum dynamics \cite{PhysRevLett.102.240603,Akila_2016} (see \cite{Bertini:2025ddr} for a recent review). This idea has been applied to a range of problems in many-body physics, including the exact dynamics of correlation functions \cite{PhysRevLett.123.210601,PhysRevB.102.174307,PhysRevLett.126.100603,PhysRevB.101.094304,PhysRevLett.133.170403}, studies of entanglement properties \cite{Bertini:2018fbz,PhysRevLett.132.120402,Gopalakrishnan:2019pip,Bertini:2019gbu,Bertini:2019wkb,PhysRevLett.125.070501,Reid:2021fsg,Zhou:2022uuw,PhysRevB.107.174311,PRXQuantum.6.010324,Pathak:2026rfk}, and the spectral form factor \cite{Bertini:2018wlu,Bertini:2020mdd,PhysRevResearch.2.043403,PhysRevX.11.021051,PhysRevResearch.6.033226,PhysRevB.111.094316}, among others. This line of work led to the development of \emph{dual-unitary} circuits, which have become an important and well-studied class of exactly tractable many-body systems. Recent extensions and generalizations of dual-unitarity have also been developed in several directions, including hierarchical/generalized dual-unitary circuits \cite{Yu:2023tmz} and generalized space-time duality \cite{Jonay:2021kgl,PhysRevResearch.6.033271,PhysRevLett.132.250402,PhysRevResearch.7.L012011}. The kicked-field Ising model (KFIM) \cite{prosenkfim,PhysRevE.65.036208,Prosen:2007hwp} is a particularly important example within this class, with additional structure that makes it especially useful. Because of its dual-unitary property \cite{Akila_2016}, the KFIM often serves as a minimal setting for studying quantum chaos and non-equilibrium dynamics in interacting many-body systems.

\begin{figure}[H]
    \centering
    \includegraphics[width=.9\linewidth]{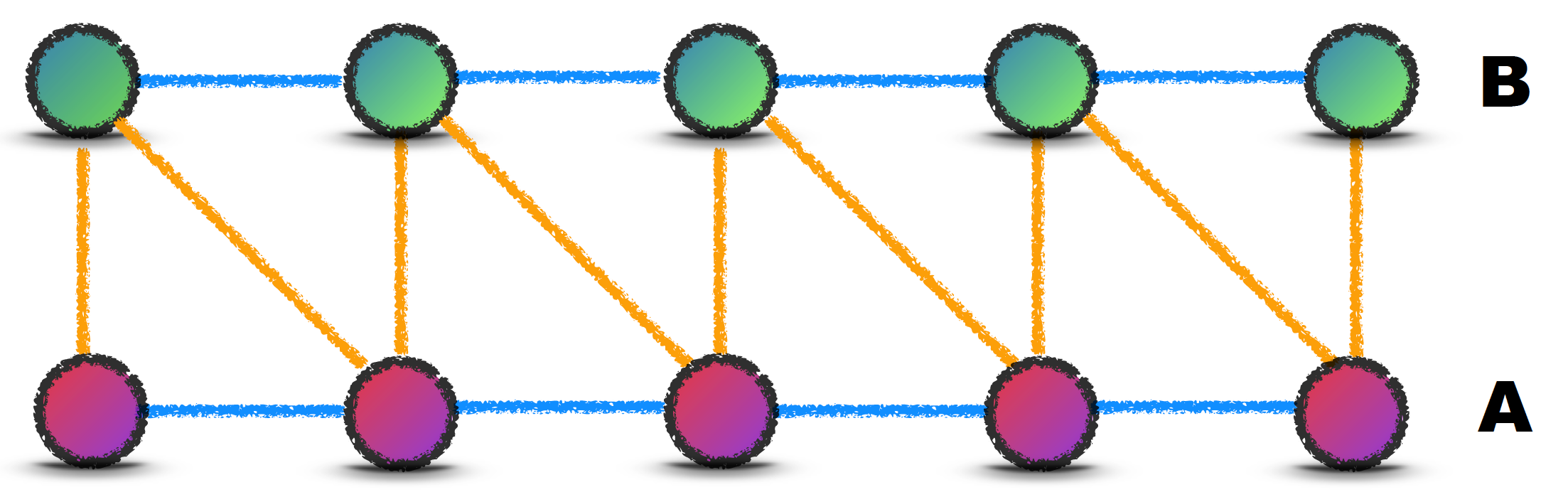}
    \caption{ Schematic of the range $r=2$ construction. Blue links form two self-dual KFIM sublattices, while orange links are arbitrary inter-sublattice couplings. These couplings make the Floquet dynamics genuinely coupled, yet they do not spoil dual-unitarity. 
    }
    \label{fig:kfimladder}
\end{figure}

Most exact results based on this framework, however, are restricted to nearest-neighbor interactions. Extending dual-unitary solvability to finite-range interactions is a natural but nontrivial step, since additional couplings typically obstruct it. This raises a natural question: \emph{Can dual-unitary solvability survive genuinely finite-range interactions?} We answer this question in the affirmative. A trivial way to obtain it is to place several independent and uncoupled dual-unitary chains side by side. Our construction, however, is different: we start from individual dual-unitary chains that are then coupled by arbitrary inter-sublattice interactions that are generic at the level of the Floquet Hamiltonian. The key observation is that, after grouping sites into enlarged unit cells, these inter-sublattice couplings enter the space-time-dual transfer problem only as internal unitary phase factors, and hence do not spoil dual-unitarity. We then study the entanglement dynamics of the model exactly for a class of solvable initial states using space-time duality. For the minimal nontrivial range $r=2$, realized as a ladder geometry (see Fig. \eqref{fig:kfimladder}) with a single diagonal interaction, we explicitly show the mechanism. Our construction demonstrates that dual-unitarity and exact entanglement dynamics can persist beyond strictly nearest-neighbor circuits, and extends naturally to more general finite-range geometries and also to the case of heterogeneous local Hilbert spaces (see \cite{supp} for a minimal realization).
\textit{Setup and results.}
We now introduce a model with range-$r$ interactions. Consider a spin-$1/2$ chain of length $L$, with sublattices $A_{i}$; $i = 1,2, \cdots,r$. 
The sublattices composed of sites with fixed $k \bmod r\, \forall \,k \in\{0, 1, \cdots , r-1\}$ are dual-unitary.  The Floquet unitary is given by 
\begin{align}\label{suppeq:kfitri}
    U & = e^{-iH_{x}}e^{-i(\sum_{k=0}^{r-1}H_{z}^{(k \bmod r)}+H_{z}^{AB}+H_{z})}, \\
\text{where{, }}    H_x &= \sum_{k=0}^{r-1}( b_{(k \bmod r)} \sum_{n \in {(k \bmod r)}}\sigma^{x}_{n}) , \nonumber\\
    H_{z}^{(k \bmod r)} &= \sum_{k=0}^{r-1}( J_{(k \bmod r)} \sum_{n\in (k \bmod r)}\sigma^{z}_{n}\sigma^{z}_{n+r}), \nonumber \\
    H_{z}^{AB} &=  \sum_{k=1}^{r-1}\sum_{n}J_{n,k}\sigma^{z}_{n}\sigma^{z}_{n+k}, \nonumber \\
    H_{z}&= \sum_{n}h_{n}\sigma_{n}^{z}
\end{align}
We impose periodic boundary conditions and choose $L$ to be a multiple of $r$ to keep a symmetric structure, such as in the case $r=2$ as shown in Fig. \eqref{fig:kfimladder}.

To present the model and the mechanism explicitly, we use the minimal nontrivial case of range $r= 2$ interactions. The results generalize because there are no extra assumptions required for general $r$. The Floquet unitary, denoted by $U$, is given by 
\begin{align}\label{eq:kfilr}
    U &= e^{-iH_{K}}e^{-i H_{I}}, \\
\text{where{, }}    
H_{I}&= H_{z}^{AA}+H_{z}^{BB}+H_{z}^{AB}+H_{z} \nonumber \\
H_K &= b_{A} \sum_{n \in A}\sigma^{x}_{n} + b_{B}\sum_{n \in B}\sigma^{x}_{n}, \nonumber\\
    H_{z}^{AA} &= J_{A} \sum_{n\in A}\sigma^{z}_{n}\sigma^{z}_{n+2}, \nonumber \\
    H_{z}^{BB} &= J_{B} \sum_{n\in B}\sigma^{z}_{n}\sigma^{z}_{n+2}, \nonumber \\
    H_{z}^{AB} &=  \sum_{n}J_{n}\sigma^{z}_{n}\sigma^{z}_{n+1}, \nonumber \\
    H_{z}&= \sum_{n}h_{n}\sigma_{n}^{z}\nonumber
\end{align}
and we have periodic boundary conditions. The constrained sum is understood as: $\sum_{n\in A}\sigma_{n}\sigma_{n+2}= \sigma_{1}\sigma_{3}+\sigma_{3}\sigma_{5}+\cdots$ and $\sum_{n\in B}\sigma_{n}\sigma_{n+2}= \sigma_{2}\sigma_{4}+\sigma_{4}\sigma_{6}+\cdots$. Similarly, the unconstrained sum is: $\sum_{n}\sigma_{n}\sigma_{n+1}= \sigma_{1}\sigma_{2}+\sigma_{2}\sigma_{3}+\cdots$.
It is useful to regard $(2j-1,2j)$ as an enlarged unit cell. Using this representation, for generic $J_{n}$,$ h_{n}$ and 
$|J_{A}|=|J_{B}|= |b_{A}|=|b_{B}|=\frac{\pi}{4}$ , the model has dual unitarity. It is also straightforward to generalize it to range $r>2$ interactions such that the sublattices: $\{1,1+r,1+2r,\cdots,L-r\}$, $\{2,2+r,2+2r,\cdots,L-r+1\},$ $\cdots,$ $\{r,2r,3r,...,L\}$ are individually dual-unitary. The model is non-integrable for generic values of $J_{n}, h_{n}$. We verify it using quasi energy level statistics and spectral-form-factor, which is found in agreement with circular orthogonal matrix (COE) distribution \cite{supp}.

For the present analysis, however, we restrict to the minimal case of $r=2$, given by Eq. \eqref{eq:kfilr}. 
As a starting point we consider the partition function $\tr\left[U^{t} \right]$. We obtain the following trace duality property

\begin{prop}\label{pro:1}
    The model defined using Eq. \eqref{eq:kfilr} can be shown to satisfy
    \begin{equation}
        \tr\left[U^{t} \right]= \tr\left[\tilde{U}^{L/2} \right] 
    \end{equation} 
where $\tilde{U}= e^{-i \tilde{H}_{K}t} e^{-i \tilde{H}_{I}t}$ is the dual evolution operator, with enlarged unit cell. The dual Hamiltonians $\tilde{H}_{K}$ and $\tilde{H}_{I}$ are defined as 
\begin{align}
    \tilde{H}_{I}&= \tilde{J}(b) \sum_{\tau=1}^{t}(\sigma^{z,A}_{\tau}\sigma^{z,A}_{\tau+1} + \sigma^{z,B}_{\tau}\sigma^{z,B}_{\tau+1})+ \sum_{\tau=1}^{t}J_{\tau}\sigma_{\tau}^{z,A}\sigma_{\tau}^{z,B}\nonumber \\
    &+\sum_{\tau=1}^{t}(h^{A}_{\tau}\sigma_{\tau}^{z,A}+h^{B}_{\tau}\sigma_{\tau}^{z,B}),  \\
    \tilde{H}_{K}&= \sum_{\tau=1}^{t}(\tilde{b}(J_{A})\sigma_{\tau}^{x,A}+\tilde{b}(J_{B})\sigma_{\tau}^{x,B}). 
\end{align}
The trace duality holds in general. At the special point $|J_{A}|=|J_{B}|= |b_{A}|=|b_{B}|=\frac{\pi}{4}$ the model is dual-unitary. For range-$r$ interactions, Property \eqref{pro:1} can be shown to be true by taking $r$ sites as a single enlarged unit cell. 
\end{prop}
See \cite{supp} for a proof. This shows that the diagonal interaction as shown in ladder configuration Fig. \eqref{fig:kfimladder} acts as a unitary \emph{phase} factor in the enlarged unit cell and does not spoil dual-unitarity of the model. 

Now, we focus on the spread of entanglement for a given initial state. We consider the initial state, which is a product state, as follows:
\begin{equation}\label{eq:inistate}
    \ket{\psi_{\theta,\phi}}= \bigotimes_{k=1}^{L} \,\left(\cos\left(\frac{\theta_{k}}{2}\right) \ket{\uparrow} + e^{i\phi_{k}} \sin\left(\frac{\theta_{k}}{2}\right) \ket{\downarrow}\right).
\end{equation}
Furthermore, we specifically consider the two classes of states: transverse ($\mathcal{T}$) and longitudinal ($\mathcal{L}$), with following parameters
\begin{align}
    &\mathcal{T}= \{\ket{\psi_{\pi/2,\phi}} \forall\, \theta_{k}= \frac{\pi}{2}, \phi \in [0,2\pi] \},  \\
    &\mathcal{L}= \{\ket{\psi_{\bar{\theta},\phi}} \bar{\theta}=\{ 0,\pi\}, \phi_{k} \in [0,2\pi]\}.
\end{align}
The above states are called solvable in the sense that the dynamics can be calculated exactly if the initial state belongs to either $\mathcal{T}$ or $\mathcal{L}$ class \cite{Bertini:2018fbz}. The results obtained for $\mathcal{L}$ states are the same as $\mathcal{T}$ states, delayed by a period. The other states which do not belong to these classes will be called \emph{generic}.

The entanglement is quantified using the reduced density matrix defined as $\rho_{X}(t) = \tr_{\bar{X}}(\ket{\psi_{\theta,\phi}(t)}\bra{\psi_{\theta,\phi}(t)})$, for a bipartition $X \otimes \bar{X}$ as the total system. The traced-out subsystem $\bar{X}$ consists of $N$ sites and subsystem $X$ consists of the remaining $L-N$ sites. Using the initial state, Eq. \eqref{eq:inistate}, we finally obtain 
\begin{equation}\label{eq:rhotr}
    \tr[(\rho_{X}(t))^{n}]= \tr\left[ \left( \prod_{j=1}^{N} \mathbb{T}[\theta_{j},\phi_{j}]\right) \left(\prod_{j=N+1}^{L} \mathbb{R}[\theta_{j},\phi_{j}]\right) \right],
\end{equation}
where $\mathbb{T}^{(\nu)}[\theta,\phi]$ is a dual transfer operator that acts non-trivially only on the $\nu$-th copy and is explicitly given as product of a Hermitian, a projection and a unitary operator.
\begin{align}
    \mathbb{T}^{(\nu)}[\theta,\phi]= \mathbb{H}^{z}_{\nu,1}.\mathbb{G}^{z}_{\nu,t}.\mathbb{U}^{(\nu)}[h].
\end{align} 
The Hermitian operator $\mathbb{H}^{z}_{\nu,1}$ and the projector  $\mathbb{G}^{z}_{\nu,t}$ can be shown \cite{supp} to factor into contributions from sublattices $A$ and $B$. The operator $\mathbb{U}^{(\nu)}[h]$ contains a non-factorizing part which contributes a unitary phase. The operator $\mathbb{R}[\theta,\phi]$ is same as $\mathbb{T}[\theta,\phi]$ upto a cyclic permutation \cite{supp}. Using these, we can explicitly show the following property holds-
\begin{prop}\label{eq:prop2}
    Let $\textrm{Spec}(\mathbb{T})$ denote the spectrum of $\mathbb{T}$, then we can show the following
    \begin{enumerate}
        \item $|\lambda_{i}| \leq \lambda_{max}^{(AB)} \equiv (1+|\cos(\theta)|)^{2n}=\lambda_{max}^{(A)}\lambda_{max}^{(B)}\,\,\forall{} \lambda_{i}\in\textrm{Spec}(\mathbb{T})$
        \item For an eigenvalue $\lambda$ of $\mathbb{T}$ that satisfies $|\lambda|= \lambda_{\max}$ we have
        \begin{enumerate}
            \item[(a)] The geometric and algebraic multiplicities of $\lambda$ coincide i.e. it has trivial Jordan blocks. 
            \item[(b)] the left eigenvector $\bra{A}$ then satisfies
            $$\bra{A}\prod_{\nu=1}^{n}\mathbb{H}_{\nu,1}^{z} = \lambda_{max} \bra{A},$$
            $$\bra{A}\prod_{\nu=1}^{n}\mathbb{G}^{z}_{\nu,1}= \bra{A},$$
            $$\bra{A}\prod_{\nu=1}^{n}\mathbb{U}= e^{i\alpha}\bra{A}.$$
        \end{enumerate}
 \end{enumerate}
\end{prop}
See \cite{supp} for a proof. An important consequence of these properties is that the properties of the $r=2$ model can be fully determined by the dual-unitary properties of the two sublattice dual-unitary chains.

\begin{figure}[H]
    \centering
    \includegraphics[width=\linewidth]{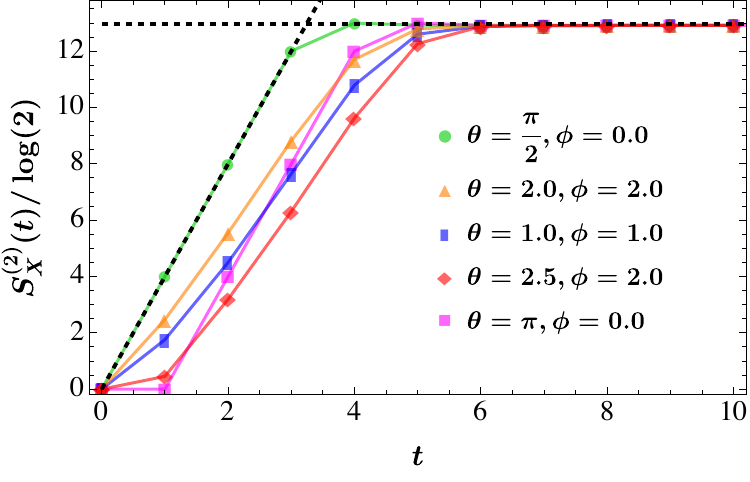}
    \caption{ The R\'enyi-2 entropy for range-$2$ model evolving from various initial states. We choose total spins $L= 30$ and subsystem size to be $N=13$. The initial state is of the form defined by Eq. \eqref{eq:inistate} and $\theta_{k}=\theta$, $\phi_{k}=\phi$. The black dashed slant line denotes the maximal growth $4t$, while the horizontal dashed line denotes the maximal entropy $N \log2$.}
    \label{fig:finitel}
\end{figure}

Finally, to quantify the amount of entanglement we evaluate the $\alpha-$R\'enyi entropies which are defined as 
\begin{equation}
    S^{(\alpha)}_{X}(t)= \frac{1}{1-\alpha}  \ln(\tr\left[(\rho_{X}(t))^{\alpha}) \right] \quad \alpha > 0.
\end{equation}
Using Eq. \eqref{eq:rhotr} and property \ref{eq:prop2}, we thus obtain that the R\'enyi entropy for initial transverse state is given by
\begin{align} \label{eqn:entfinal}
    S_{X}^{(\alpha)}(t)&=  S_{X}^{(\alpha,A)}(t)+S_{X}^{(\alpha,B)}(t) = \min(4t,N) \ln(2).
\end{align}
Thus the range $r=2$ model produces entanglement at twice the rate of the nearest-neighbor self-dual KFIM, while remaining exactly solvable.
For a general case of range-$r$ interactions, similar mechanism can then be used to show that for range-$r$ interaction we obtain 
\begin{align} \label{eqn:entfinalr}
    S_{X}^{(\alpha)}(t)&=  \sum_{k=1}^{r}S_{X}^{(\alpha,A_{i})}= \min(2rt,N) \ln(2),
\end{align}
where each $A_{i}$ denotes the sublattice that is dual-unitary. The interaction range therefore acts as a tunable multiplier of the exact entanglement production rate.
Using Eq. \eqref{eqn:entfinalr} we can deduce that the spectrum of the reduced density matrix is flat
\begin{prop}\label{eq:prop3}
    The spectrum of the reduced density matrix is flat and is given by 
    \begin{equation}
        \textrm{Spec}[\rho_{X}(t)]= \{2^{-\min(2rt,N)},0\},
    \end{equation}
    with degeneracy of nonzero eigenvalue equal to $2^{\min(2rt,N)}$.
\end{prop}
This then implies that all the $n-$R\'enyi entropies are the same, independent of the value of $n$.
We further note that the result given by Eq. \eqref{eqn:entfinalr} holds in general for the case when the initial state chosen is not uniform over the whole chain in which case $S^{(\alpha),A_{i}}_{X}(t)$ is entanglement, which is not necessarily the same. A minimal example of this case would be, $\mathcal{T}$ states in sublattice $A_{1}$ and $\mathcal{L}$ states in sublattice $A_{2}$. The total entanglement entropy will thus be given by 
\begin{equation}
    S^{(\alpha)}_{X}(t)= \min(2t,N_{A_{1}})\ln(2)+  \min(2\max(t-1,0),N_{A_{2}})\ln(2),
\end{equation}
where $N_{A_{1}}, N_{A_{2}}$ are the numbers of sites in sublattices $A_{1}$ and $A_{2}$ respectively after partial tracing. 
To further understand the dynamics of generic initial states we study the model numerically in  Fig. \eqref{fig:finitel} with $L=30$ and $N=13$.

\begin{figure}[H]
    \centering
    \includegraphics[width=\columnwidth]{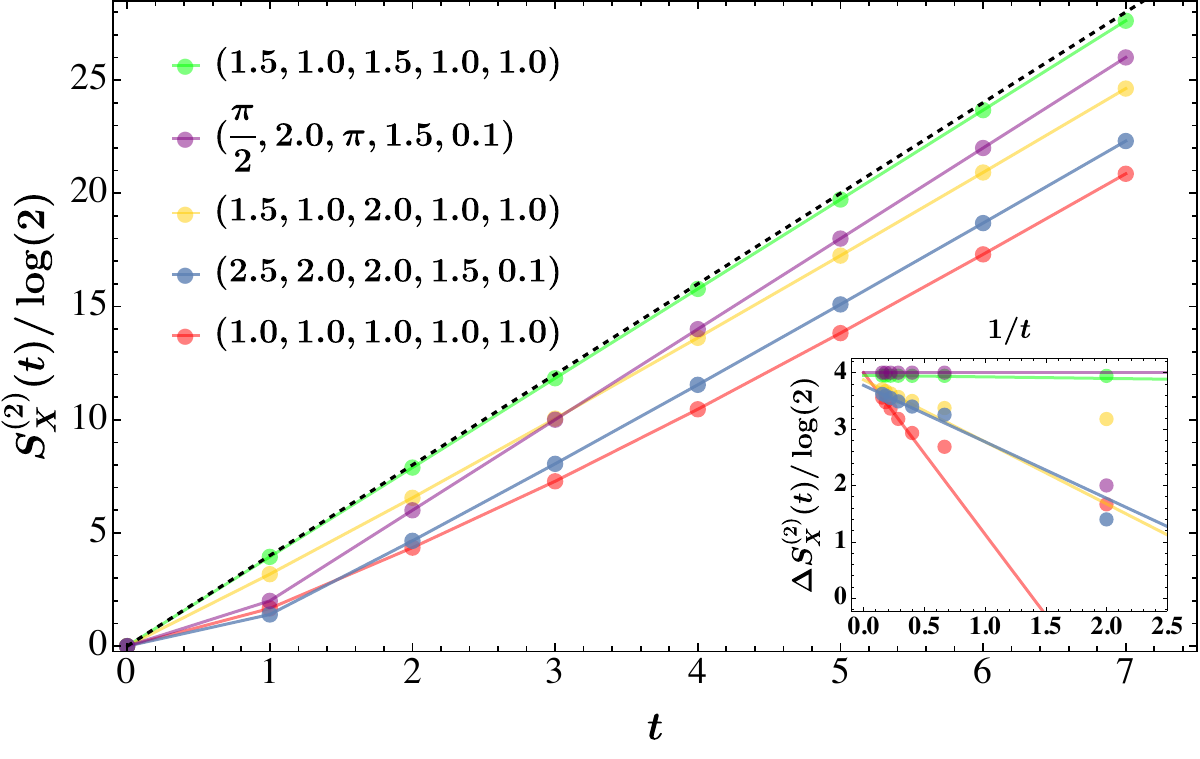}
    \caption{The R\'enyi-2 entropy for a range-$2$ model from various initial states, in the thermodynamic limit. The parameter values in the legend correspond to the tuple $(\theta_{A},\phi_{A},\theta_{B},\phi_{B},h_{j})$, where $\theta_{A/B},\phi_{A/B}$ denote the uniform initial state parameters on sublattice $A/B$. The dashed black line denotes the maximal slope of $4t$, obtained for solvable states. The inset shows the instantaneous slope, $\frac{\Delta S_{X}^{(2)}(t-1/2)}{\log(2)}$ as a function of $1/t$. The points are obtained using the duality mapping and the straight lines are obtained by linearly extrapolating the last three points.}
    \label{fig:renyi2dual}
\end{figure}

The results for the solvable states are in agreement with the analytical result \eqref{eqn:entfinal}. 
To further probe the dynamics in the thermodynamic limit, $L \rightarrow \infty, N \rightarrow \infty$, we use the duality mapping (see \cite{supp} for details on the numerics). The results for R\'enyi-2 are shown in Fig. \eqref{fig:renyi2dual}. We observe deviations from the asymptotic slope at the early times accessible numerically. However, we can obtain a more refined analysis using the instantaneous slope, which is defined as follows
\begin{equation}
    \Delta S^{(\alpha)}_{X}\left(t-\frac{1}{2}\right)= S^{(\alpha)}_{X}(t)- S^{(\alpha)}_{X}(t-1).
\end{equation}

Numerically, we observe that the instantaneous slope becomes a linear function of $1/t$ at late times, shown in Fig.\eqref{fig:renyi2dual} (inset). Further extrapolating the results we obtain the behavior in the $t \rightarrow \infty$ limit. We obtain that the results are consistent with 

\begin{equation}
 \lim_{t \rightarrow \infty}\frac{\Delta S^{(\alpha)}_{X}\left(t\right)}{\ln(2)} = 4.
\end{equation}

\textit{Discussion.} In this work, we introduce a class of Floquet models with finite-range interactions. Using the space-time duality we then obtained exact entanglement dynamics for a solvable class of initial states in these models. The basic idea is to couple sublattices through range-$r$ interactions, while keeping each sublattice individually dual-unitary. The resulting model is no longer nearest-neighbor in the original lattice; nevertheless, the inter-sublattice interactions preserve dual-unitarity because they enter the space-time-dual transfer problem as internal unitary phases of an enlarged cell. This provides a simple way of extending dual-unitary solvability beyond strictly nearest-neighbor architectures. For the minimal nontrivial case $r=2$, we showed explicitly that the entanglement growth rate is twice that of the usual dual-unitary KFIM. More generally, the same reasoning suggests that range-$r$ interactions enhance the growth rate by a factor of $r$ (the velocity still being unity), in terms of the original lattice degrees of freedom. Furthermore, for the $r=2$ model, the level statistics and spectral form factor indicate that the coupled system has chaotic spectral properties, showing that the exact solvability described here is not a consequence of integrability. If $r$ is allowed to scale with system size, the same construction suggests that a product state can reach volume-law entanglement in $\mathcal{O}(1)$ Floquet steps. The framework suggests several interesting directions beyond what is discussed here. In particular, it can be extended to sublattices with different local Hilbert-space dimensions. For example, one may couple a kicked Ising sublattice with local dimension $d=2$ to a kicked Potts sublattice with local dimension $d=3$ \cite{PhysRevB.105.144306,Claeys:2024tuy} (also see \cite{supp}). The entanglement dynamics remains analytically tractable in this setting and is obtained from the combined contributions of the two sublattices. More broadly, this suggests that the current framework provides a route to exact solvability in more general settings, where interactions need not be strictly nearest-neighbor and local Hilbert-space dimensions are heterogeneous. 

\emph{Acknowledgments:}
The author is grateful to Bruno Bertini for various helpful discussions. The author also acknowledges Abhik for a careful reading of the draft.  
The work is supported by JST CREST (Grant No. JPMJCR24I2). A part of numerical calculations have been performed using the computational facilities of the Yukawa Institute for Theoretical Physics. 

\emph{Note:} During the completion of this work author became aware of the independent study of local correlations in a related model \cite{Osipov:2026xyr}. Author acknowledges Boris Gutkin for the correspondence on the same.  
\twocolumngrid
\bibliography{references}
\end{document}


\title{Supplemental Materials: Exact Entanglement Dynamics in Finite-Range Floquet
Systems via Space-Time Duality\\
}
\author{Tanay Pathak\,\,\href{https://orcid.org/0000-0003-0419-2583}
{\includegraphics[scale=0.05]{orcidid.pdf}}}
\email{pathak.tanay.4s@kyoto-u.ac.jp}
\affiliation{Department of Physics, Kyoto University, Kitashirakawa Oiwakecho, Sakyo-ku, Kyoto 606-8502, Japan}
\maketitle
In this supplemental material, we provide the proofs of various properties of the main text along with other supplemental information. In particular:
\begin{itemize}
\item In Section \ref{suppsec:tracedual} we show the duality of the trace for the range-$2$ KFIM and prove Property 1. 
\item In Section \ref{suppsec:spectral} we study some spectral properties of the range-$2$ model. 
    \item In Section \ref{sec:replica} we set up the replica trick for the reduced density matrix. 
    \item In Section \ref{suppsec:propproof} we prove Property 2 of the main text. 
    \item In Section \ref{suppsec:duality} we give details of numerics in the thermodynamic limit using duality mapping.
    \item In Section \ref{suppsec:d2d3} we give a concrete realization of a model with sublattices having local dimensions $2$ and $3$.
\end{itemize}
\section{Duality of traces and proof of Property 1 }\label{suppsec:tracedual}
\begin{figure}[H]
    \centering
    \includegraphics[width=0.6\linewidth]{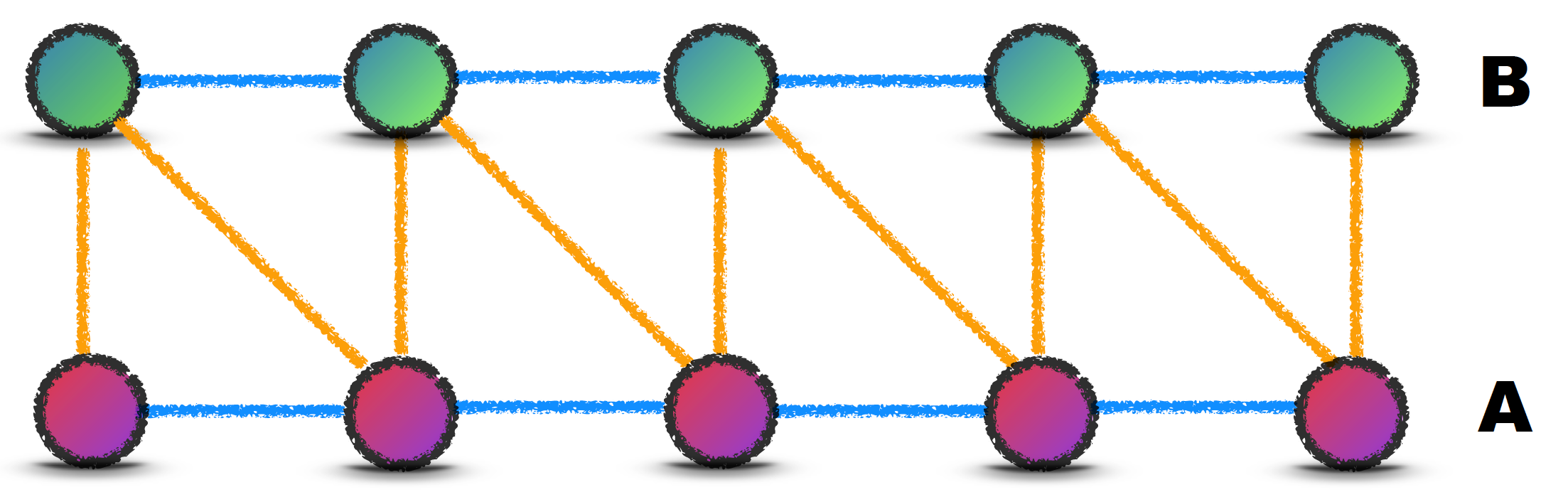}
    \caption{ Schematic of the range $r=2$ construction. Black links form two self-dual KFIM sublattices, while gray links are arbitrary diagonal inter-sublattice couplings. These couplings make the Floquet dynamics genuinely coupled, yet they do not spoil dual-unitarity in the enlarged unit cell.}
    \label{sfig:kfimladder}
\end{figure}

We first consider the minimal model of range $r=2$ interaction, which is a spin-$1/2$ chain of length $L$, which is even (see Fig. \eqref{sfig:kfimladder}). The odd and even sites, forming sublattices $A$ and $B$ are separately dual-unitary and the two sublattices are then coupled as shown in Fig. \eqref{sfig:kfimladder}. The Floquet unitary is given by 
\begin{align}\label{suppeq:kfitri}
    U[J_{A},J_{B},h,\mathbf{J_{n}}]\equiv U & = e^{-iH_{x}}e^{-i(H_{z}^{AA}+H_{z}^{BB}+H_{z}^{AB}+H_{z})}, \\
\text{where{, }}    H_x &= b_{A} \sum_{n \in A}\sigma^{x}_{n} + b_{B}\sum_{n \in B}\sigma^{x}_{n}, \nonumber\\
    H_{z}^{AA} &= J_{A} \sum_{n\in A}\sigma^{z}_{n}\sigma^{z}_{n+2}, \nonumber \\
    H_{z}^{BB} &= J_{B} \sum_{n\in B}\sigma^{z}_{n}\sigma^{z}_{n+2}, \nonumber \\
    H_{z}^{AB} &=  \sum_{n}J_{n}\sigma^{z}_{n}\sigma^{z}_{n+1}, \nonumber \\
    H_{z}&= \sum_{n}h_{n}\sigma_{n}^{z},
\end{align}
and we have periodic boundary conditions. 
The setup can be viewed as in Fig. \eqref{sfig:kfimladder}.

Next we consider the quantity
\begin{align} 
&\tr\left[\left(U[\boldsymbol{h}]\right)^t\right]=  \sum_{\left\{s_\tau\right\}}\left\langle\boldsymbol{s}_1\right| U[\boldsymbol{h}]\left|\boldsymbol{s}_t\right\rangle\left\langle\boldsymbol{s}_t\right| U[\boldsymbol{h}]\left|\boldsymbol{s}_{t-1}\right\rangle \cdots\left\langle\boldsymbol{s}_2\right| U[\boldsymbol{h}]\left|\boldsymbol{s}_1\right\rangle \nonumber\\ 
= & \left(\frac{\sin 2 b}{2 i}\right)^{(L t) / 2}\times \\
&\sum_{\left\{s_{\tau, j}\right\}}\left\{\exp \left[-i \tilde{J}(b_{A}) \sum_{j\in A} s_{1, j} s_{t, j}-i \tilde{J}(b_{B}) \sum_{j\in B} s_{1, j} s_{t, j}-i J \sum_{j=1}^L s_{t, j} s_{t, j+1}-i \sum_{j\in A} J_{A}s_{1, j} s_{1, j+2}-i \sum_{j\in A} J_{B}s_{1, j} s_{1, j+2} \right. \right. \nonumber \\
&\left. -i \sum_{j=1}^L h_j s_{t, j}\right] \nonumber\\
&\times \exp \left[-i \tilde{J}(b_{A}) \sum_{j\in A} s_{t, j} s_{t-1, j}-i \tilde{J}(b_{B}) \sum_{j\in B} s_{t, j} s_{t-1, j}-i J \sum_{j=1}^L s_{t-1, j} s_{t-1, j+1}-i \sum_{j\in A} J_{A}s_{t-1, j} s_{t-1, j+2}\right. \nonumber \\
&\left.-i \sum_{j\in A} J_{B}s_{t-1, j} s_{t-1, j+2}-i \sum_{j=1}^L h_j s_{t-1, j}\right] \nonumber \\
& \vdots \nonumber\\
&\times \exp \left[-i \tilde{J}(b_{A}) \sum_{j\in A} s_{2, j} s_{1, j}-i \tilde{J}(b_{B}) \sum_{j\in B} s_{2, j} s_{1, j}-i J \sum_{j=1}^L s_{1, j} s_{1, j+1}-i \sum_{j\in A} J_{A}s_{1, j} s_{1, j+2}\right. \nonumber \\
&\left.-i \sum_{j\in A} J_{B}s_{1, j} s_{1, j+2}-i \sum_{j=1}^L h_j s_{1, j}\right], \nonumber \\
\end{align}
 where 
 \begin{align}
     \tilde{J}(b)= - \frac{\pi}{4}- \frac{i}{2}\ln(\tan(b)).
 \end{align}

Reorganizing the sum we obtain 
\begin{align} 
&\operatorname{tr}\left[\left(U\right)^t\right]=  \left(\frac{\sin 2 b}{2 i}\right)^{[(L t) / 2]} \nonumber \\
&\sum_{\left\{s_{\tau, j}\right\}}\left\{\exp \left[-i \tilde{J}(b_{A}) \sum_{\tau=1}^t s_{\tau, 1} s_{\tau+1,1}-i \tilde{J}(b_{B}) \sum_{\tau=1}^t s_{\tau, 2} s_{\tau+1,2}-i J \sum_{\tau=1}^t s_{\tau, 1} s_{\tau, 2}-i \sum_{\tau=1}^t J_{A} s_{\tau, 1} s_{\tau, L-1}-i \sum_{\tau=1}^t J_{B} s_{\tau, 2} s_{\tau, L} \right.\right. \nonumber \\
&\left. -i \sum_{\tau=1}^t (h_1 s_{\tau, 1}+h_2 s_{\tau, 2})\right] \nonumber\\ 
&\exp \left[-i \tilde{J}(b_{A}) \sum_{\tau=1}^t s_{\tau, 3} s_{\tau+1,3}-i \tilde{J}(b_{B}) \sum_{\tau=1}^t s_{\tau, 4} s_{\tau+1,4}-i J \sum_{\tau=1}^t s_{\tau, 3} s_{\tau, 4}-i \sum_{\tau=1}^t J_{A} s_{\tau, 3} s_{\tau, 1}-i \sum_{\tau=1}^t J_{B} s_{\tau, 4} s_{\tau, 2} \right. \nonumber \\
&\left. -i \sum_{\tau=1}^t (h_3 s_{\tau, 3}+h_4 s_{\tau, 4}\right] \nonumber\\
& 
\vdots \nonumber\\ 
&\exp \left[-i \tilde{J}(b_{A}) \sum_{\tau=1}^t s_{\tau, L-1} s_{\tau+1,L-1}-i \tilde{J}(b_{B}) \sum_{\tau=1}^t s_{\tau, L} s_{\tau+1,L}-i J \sum_{\tau=1}^t s_{\tau, L} s_{\tau, L-1}-i \sum_{\tau=1}^t J_{A} s_{\tau, L-11} s_{\tau, L-3} \right. \nonumber \\
&\left. -i \sum_{\tau=1}^t J_{B} s_{\tau, L} s_{\tau, L-2} -i \sum_{\tau=1}^t (h_L s_{\tau, L}+h_{L-1} s_{\tau, L-1})\right] \nonumber\\
&= \tr\left[ (\tilde{U})^{L/2}\right].
\end{align}

So we can introduce the dual Floquet operator as
\begin{equation}
    \tilde{U}= e^{-i\tilde{H}_{k}}e^{-i\tilde{H}_{I}},
\end{equation}

where 
\begin{align}
   \tilde{H}_{I}&= \tilde{J}(b) \sum_{\tau=1}^{t}(\sigma^{z,A}_{\tau}\sigma^{z,A}_{\tau+1} + \sigma^{z,B}_{\tau}\sigma^{z,B}_{\tau+1})+ \sum_{\tau=1}^{t}J_{\tau}\sigma_{\tau}^{z,A}\sigma_{\tau}^{z,B}+\sum_{\tau=1}^{t}(h^{A}_{\tau}\sigma_{\tau}^{z,1}+h^{B}_{\tau}\sigma_{\tau}^{z,B}),  \\
   \tilde{H}_{K}&= \sum_{\tau=1}^{t}(\tilde{b}(J_{A})\sigma_{\tau}^{x,A}+\tilde{b}(J_{B})\sigma_{\tau}^{x,B}).
\end{align}
We will further be interested in set of parameters $|J_{A}|=|J_{B}|=|b_{A}|=|b_{B}|= \frac{\pi}{4}$. It is easy to see that at this point the dual evolution is also unitary, thus implying \emph{dual unitarity} of the model. 

A similar treatment is possible for the case when the interactions have range $r$. For that case we can consider a generic model which has dual unitarity on sublattices composed of lattice sites on $k \bmod r\, \forall \,k \in\{0, 1, \cdots , r-1\}$.  The Floquet unitary is given by 
\begin{align}\label{suppeq:kfitri}
    U & = e^{-iH_{x}}e^{-i(\sum_{k=0}^{r-1}H_{z}^{(k \bmod r)}+H_{z}^{AB}+H_{z})}, \\
\text{where{, }}    H_x &= \sum_{k=0}^{r-1}( b_{(k \bmod r)} \sum_{n \in {(k \bmod r)}}\sigma^{x}_{n}) , \nonumber\\
    H_{z}^{(k \bmod r)} &= \sum_{k=0}^{r-1}( J_{(k \bmod r)} \sum_{n\in (k \bmod r)}\sigma^{z}_{n}\sigma^{z}_{n+r}), \nonumber \\
    H_{z}^{AB} &=  \sum_{k=1}^{r-1}\sum_{n}J_{n,k}\sigma^{z}_{n}\sigma^{z}_{n+k}, \nonumber \\
    H_{z}&= \sum_{n}h_{n}\sigma_{n}^{z}
\end{align}
and we have periodic boundary conditions. We will further be interested in set of parameters 
$$|J_{(k \bmod r)}|=|b_{(k \bmod r)}|= \frac{\pi}{4}.$$ Only the range-$r$ couplings have the same magnitude and all the other couplings can be generic. 
\section{Spectral properties }\label{suppsec:spectral}
In this section we study some spectral properties of the model. Since $J_{n}=0$ gives a trivial limit of two decoupled chains, we study the model with generic $J_{n}$ and $h_{n}$.

For a given $\mathcal{D}-$ dimensional unitary operator $\mathcal{U}$, the eigenvalues are phases of the form $e^{i \epsilon_{i}}$, where $\epsilon_{i}$ is the quasi energy. Arranging the quasi energies in ascending order we have 
$$ \epsilon_{1} < \epsilon_{2} < \cdots <\epsilon_{\mathcal{D}}.$$

\begin{figure}[H]
    \centering
    \includegraphics[width=\linewidth]{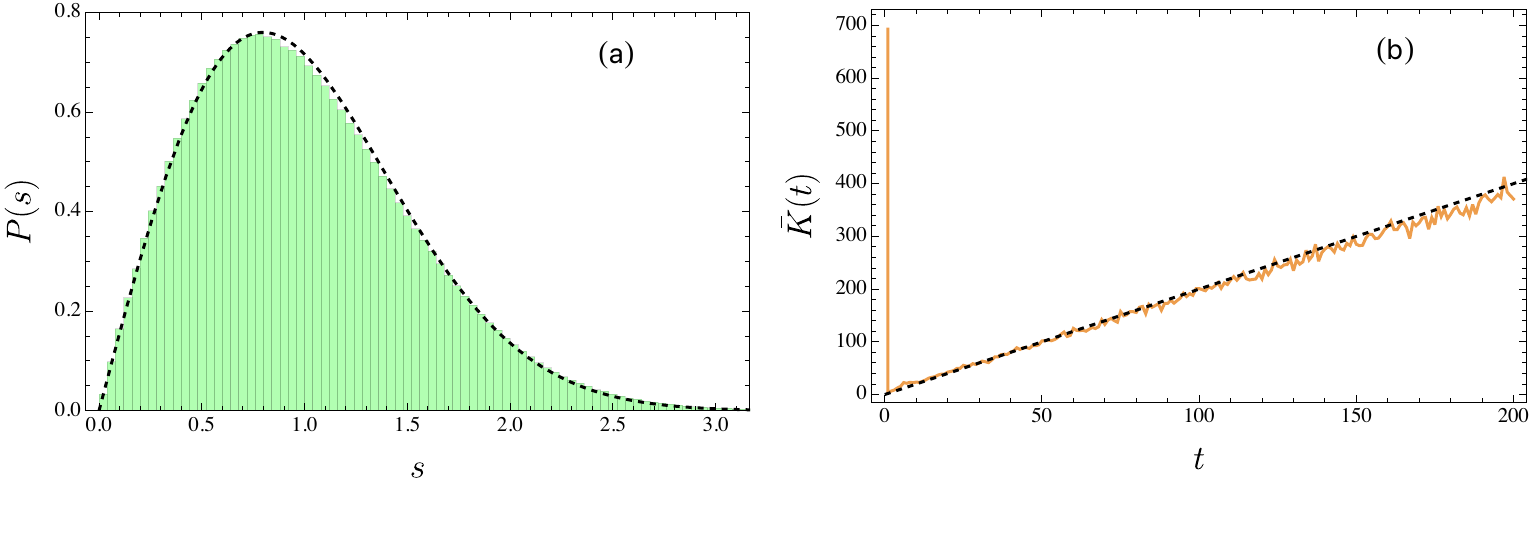}
    \caption{(a) Level spacing distribution of the (unfolded) quasi energies of the unitary operator, with $L=12$, given by Eq.\eqref{suppeq:kfitri} for generic $J_{n}$, $h_{n}$ . We choose $h_{n}=0.5$ and $J_{n}$ is chosen randomly from the interval $[0,1]$ and we average over 1024 realizations. The dashed curve corresponds to the Wigner surmise for circular orthogonal ensemble (COE). (b) Spectral form factor, as defined using Eq. \eqref{seq:sff} for the range-2 model. The result shows a clear ramp in agreement with COE results, shown as black dashed line. }
    \label{sfig:levelspec}
\end{figure}

We can then define the spacing as $s_{i}= \epsilon_{i+1}-\epsilon_{i}$. The distribution of these spacings, $P(s)$ is an indicator of chaotic or integrable behavior of the system. For an integrable system it is given by Poisson distribution while for a chaotic system it is in agreement with random matrix theory (RMT). In Fig. \eqref{sfig:levelspec} (a) we show the results for the spacing distribution of the quasi energies of the Floquet operator $U$ given by Eq.\eqref{suppeq:kfitri}. We observe that for both cases the prediction matches well with the circular orthogonal ensemble (COE) results \cite{mehta2004random}. 

Using the spacing one can also define a level spacing ratio \cite{PhysRevLett.110.084101} as $r_{i}= \frac{\min{(s_{i},s_{i+1})}}{\max{(s_{i},s_{i+1})}}$. The spacing ratios also carry the same information as the level spacing distribution with an added benefit that it does not require unfolding. Depending on whether the system is integrable (Poisson) or chaotic (RMT like), the average of the level spacing ratios has following values 
\begin{equation}
    \braket{r}= \begin{cases}
        2\ln(2)-1, \quad \text{Integrable},\\
        0.5307, \quad \text{GOE/COE}.
    \end{cases}
\end{equation}
For our case we found $\braket{r}= 0.5307 \pm 0.00046$ in agreement with the above RMT predictions. 

We can further study the correlation among quasi energies using spectral form factor which is defined as 
\begin{equation}\label{seq:sff}
    \bar{K}(t)= |\tr(U^{t})|^{2}
\end{equation}
where the bar denotes the averaging over disorder realizations. For a typical chaotic system this quantity is known to show a universal ramp and plateau behavior. We study this quantity for the present model, as shown in Fig. \eqref{sfig:levelspec} (b), and find agreement with the COE prediction $\bar{K}(t)_{\text{COE}} = 2t$, confirming the chaotic nature of the model.
\section{Entanglement entropy}\label{sec:replica}

Now, we move on to the evaluation of entanglement entropy. For that we will be interested in the following class of initial states.

\begin{equation}
    \ket{\psi_{\theta,\phi}}= \bigotimes_{k=1}^{N} \,\left(\cos\left(\frac{\theta_{k}}{2}\right) \ket{\uparrow} + e^{i\phi_{k}} \sin\left(\frac{\theta_{k}}{2}\right) \ket{\downarrow}\right) .
\end{equation}
Furthermore, there are two classes of states which are of interest \cite{Bertini:2018fbz}: Transverse ($\mathcal{T}$) and Longitudinal ($\mathcal{L}$), which are given by following parameters
\begin{align}
    &\mathcal{T}= \{\ket{\psi_{\theta_{k},\phi}} \forall\, \theta_{k}= \frac{\pi}{2}, \phi \in [0,2\pi] \},\label{suppeq:tclass} \\
    &\mathcal{L}= \{\ket{\psi_{\bar{\theta},\phi}} \bar{\theta}=\{ 0,\pi\}, \phi_{k} \in [0,2\pi]\}. \label{suppeq:lclass}
\end{align}
Consider a bipartition $X$ and $\bar{X}$ of the time evolved state $\psi_{\theta,\phi}$. The entanglement is quantified using the density matrix of the subsystems 
\begin{align}\label{suppeq:rhoreda}
    \rho_{X}(t) &= \tr_{\bar{X}}(\ket{\psi_{\theta,\phi}(t)}\bra{\psi_{\theta,\phi}(t)}) \nonumber \\
    &= \sum_{i_{1},i_{2} \atop i'_{1},i'_{2} } \braket{i_{1},i_{2}| U[\mathbf{J},\mathbf{h}]^{t}|\psi_{\theta,\phi}} \braket{\psi_{\theta,\phi}| U[\mathbf{J},\mathbf{h}]^{-t}|i'_{1},i'_{2}} \times \ket{i_{1},i_{2}}\bra{i'_{1},i'_{2}}.
\end{align}
For brevity from now on we write $U[\mathbf{J},\mathbf{h}] \equiv U$.
The entanglement content of $\rho_{X}(t)$ is then quantified using R\'enyi entropy defined as
\begin{equation}
    S^{(n)}_{X}(t)= \frac{1}{1-n}\ln(\tr(\rho_{X}(t)^{n})), \quad n>0.
\end{equation}
So the main object that we need to evaluate is the moments $\tr(\rho_{X}(t)^{n})$. 

First note that using Eq. \eqref{suppeq:rhoreda} we have
\begin{align}\label{suppeq:trrhoa}
    \tr[(\rho_{X}(t))^{n}]  &=  \sum_{\{\mbf{a}_{i}\},\{\mbf{b}_{i}\}} \braket{\mbf{a}_{1},\mbf{b}
    _{1}| U^{t}|\psi_{\theta,\phi}} \braket{\psi_{\theta,\phi}| U^{-t}|\mbf{a}_{1},\mbf{b}_{2}} \nonumber \\
    & \times \braket{\mbf{a}_{2},\mbf{b}_{2}| U^{t}|\psi_{\theta,\phi}} \braket{\psi_{\theta,\phi}| U^{-t}|\mbf{a}_{2},\mbf{b}_{3}} \nonumber \\
    \vdots \nonumber \\
    &\times \braket{\mbf{a}_{n},\mbf{b}_{n}| U^{t}|\psi_{\theta,\phi}} \braket{\psi_{\theta,\phi}| U^{-t}|\mbf{a}_{n},\mbf{b}_{1}}.
\end{align}

To evaluate the above expression we need to evaluate the following

\begin{align}\label{suppeq:partfunc}
    \bra{ \mathbf{a},\mathbf{b}} (U)^{t})|\psi_{\theta,\phi} \rangle &= 
     \sum_{\{s_{\tau}\}} \left(\prod_{\tau=1}^{t-1} \braket{s_{\tau+1}|U|s_{\tau}}\right) \bra{\mbf{a},\mbf{b}}U\ket{s_{t}}  \bra{s_{1}}\psi_{\theta,\phi}\rangle.
\end{align}
Using the following identities
\begin{align}
    \langle \mathbf{s} | \psi_{\theta,\phi}  \rangle &= \prod_{k=1}^{L}[\cos(\theta_{k}/2)\delta_{s_{k},1}+\sin(\theta_{k}/2)e^{i\phi_{k}}\delta_{s_{k},-1}], \\
    \braket{\mathbf{s} | e^{-i(b)\sigma^{x}} | \mbf{r} } &= \sqrt{\frac{\sin(2b)}{2i}}\exp\left[ -i \tilde{J} s r\right], \quad s,r \in\{\pm 1\}, \quad \tilde{J}= \frac{\pi}{4}+ \frac{i}{2}\ln(\tan(b)).
\end{align}
we have 
\begin{align}
    \braket{\mathbf{s} | U | \mathbf{r} } &=\left(-\frac{i}{2}\right)^{L/2} \exp\left( -i\frac{\pi}{4}\sum_{n=1}^{L}s_{n}r_{n} \right) \times  \exp\left( -i \frac{\pi}{4}\sum_{n\in A}r_{n}r_{n+2} -i \frac{\pi}{4}\sum_{n\in B}r_{n}r_{n+2}- i \sum_{n=1}^{L} J_{n} r_{n}r_{n+1} - i \sum_{n}h_{n}r_{n} \right).
\end{align}

Putting together with Eq. \eqref{suppeq:partfunc} we finally obtain
\begin{align}
  &\bra{ \mathbf{a},\mathbf{b}} (U [\mathbf{h}])^{t})|\psi_{\theta,\phi} \rangle \nonumber \\  &= \sum_{\{s_{\tau,j}\}}\left(\frac{i}{2}\right)^{Lt/2} 
  \times \exp\left[-i \frac{\pi}{4}\left(\sum_{\tau =1}^{t} \sum_{j \in A} s_{\tau,j}s_{\tau,j+2} + \sum_{\tau=1}^{t}\sum_{j\in B}s_{ \tau,j}s_{\tau,j+2} +\sum_{\tau=1}^{t-1}\sum_{j=1}^{L} s_{\tau,j}s_{\tau+1,j} \right)-i\sum_{\tau=1}^{t}\sum_{j=1}^{L} J_{n}s_{\tau,j}s_{\tau,j+1}  \right] \nonumber \\
  & \times \exp\left(-i\sum_{\tau=1}^{t}\sum_{j=1}^{L} h_{n}s_{\tau,j} -i \frac{\pi}{4}\sum_{j=1}^{N}s_{t,j}a_{j} -i \frac{\pi}{4}\sum_{j=N+1}^{L}s_{t,j}b_{j-N} \right) \nonumber \\
  & \times \prod_{j=1}^{L}[\cos(\theta_{1,j}/2)\delta_{s_{1,j},1}+\sin(\theta_{1,j}/2)e^{i\phi_{1,j}}\delta_{s_{1,j},-1}].
\end{align}

Combining above with Eq. \eqref{suppeq:trrhoa} we then obtain
\begin{align}
   & \tr[(\rho_{X}(t))^{n}] = \frac{1}{2^{nLt}} \sum_{\{s_{\tau,j}\}}\nonumber\\
  &\times \exp\left[-i \sum_{\nu=1}^{2n}\mathrm{sgn}(n-\nu)\left(\sum_{j=1}^{L}\sum_{\tau =1}^{t} \frac{\pi}{4} s_{\nu,\tau,j}s_{\nu,\tau,j+2} +\sum_{j=1}^{L}\sum_{\tau =1}^{t}\left( J_{j} s_{\nu,\tau,j}s_{\nu,\tau,j+1}+h_{j}s_{\nu,\tau,j} \right)\right)  \right] \nonumber \\
  &\times \exp\left(-i\frac{\pi}{4}\sum_{\nu=1}^{2n}\sum_{\tau=1}^{t-1}\sum_{j=1}^{L} \mathrm{sgn}(n-\nu) s_{\nu,\tau,j}s_{\nu,\tau+1,j} \right) \prod_{\nu=1}^{n}\left(\prod_{j=1}^{N}(1+s_{\nu,t,j}s_{\nu+n,t,j}) \prod_{j=N+1}^{L}(1+s_{\nu,t,j}s_{n+1+\mathrm{mod}(\nu-2,n),t,j}) \right) \nonumber \\
  & \times  \prod_{\nu=1}^{2n} \prod_{j=1}^{L}[\cos(\theta_{j}/2)\delta_{s_{\nu,1,j},1}+\sin(\theta_{j}/2)e^{i\phi_{j}\mathrm{sgn}(n-\nu)}\delta_{s_{\nu,1,j},-1}].
\end{align}

We then define the following operators

\begin{align} 
&\langle \{s^{(A)}_{\nu,\tau,j}\},\{s^{(B)}_{\nu,\tau,j'}\} | \mathbb{T}_{\theta,\phi}[h] | \{s^{(A)}_{\nu,\tau,j+2}\},\{s^{(B)}_{\nu,\tau,j'+2}\} \rangle= \nonumber \\
&  \frac{1}{2^{(t-1)n}} \exp \left[ -i \sum_{\nu=1}^{2n} \mathrm{sgn}(n-\nu) \left( \sum_{\tau=1}^{t} \left( \frac{\pi}{4}( s^{(A)}_{\nu,\tau,j}s^{(A)}_{\nu,\tau,j+2}+s^{(B)}_{\nu,\tau,j'}s^{(B)}_{\nu,\tau,j'+2}) + (h^{(A)}_j s_{\nu,\tau,j}+h^{(B)}_{j'} s_{\nu,\tau,j'}) \right) \right. \right. \nonumber \\
&\left. \left. + \sum_{\tau=1}^{t} J_j s^{(A)}_{\nu,\tau,j}s^{(B)}_{\nu,\tau,j+1} + \sum_{\tau=1}^{t-1} \frac{\pi}{4} (s^{(A)}_{\nu,\tau,j}s^{(A)}_{\nu,\tau+1,j}+s^{(B)}_{\nu,\tau,j'}s^{(B)}_{\nu,\tau+1,j'}) \right) \right] \nonumber \\ 
& \times \prod_{\nu=1}^{n} \left( \frac{1+s^{(A)}_{\nu,t,j}s^{(A)}_{\nu+n,t,j}}{2} \right)\left( \frac{1+s^{(B)}_{\nu,t,j}s^{(B)}_{\nu+n,t,j}}{2} \right) \nonumber\\
&\times \prod_{\nu=1}^{2n} \left[ \cos(\theta^{(A)}_{j}/2)\delta_{s_{\nu,1,j},1} + \sin(\theta^{(A)}_{j}/2)e^{i\phi^{(A)}_{j}\mathrm{sgn}(n-\nu)}\delta_{s_{\nu,1,j},-1} \right]\left[ \cos(\theta^{(B)}_{j'}/2)\delta_{s_{\nu,1,j'},1} + \sin(\theta^{(B)}_{j'}/2)e^{i\phi^{(B)}_{j'}\mathrm{sgn}(n-\nu)}\delta_{s_{\nu,1,j'},-1} \right] \label{eq:tmatelement} \\
&\langle \{s_{\nu,\tau,j}\},\{s_{\nu,\tau,j+1}\} | \mathbb{R}_{\theta,\phi}[h] | \{s_{\nu,\tau,j+1}\},\{s_{\nu,\tau,j+2}\} \rangle= \nonumber \\
&  \frac{1}{2^{(t-1)n}} \exp \left[ -i \sum_{\nu=1}^{2n} \mathrm{sgn}(n-\nu) \left( \sum_{\tau=1}^{t} \left( \frac{\pi}{4}( s^{(A)}_{\nu,\tau,j}s^{(A)}_{\nu,\tau,j+2}+s^{(B)}_{\nu,\tau,j'}s^{(B)}_{\nu,\tau,j'+2}) + (h^{(A)}_j s_{\nu,\tau,j}+h^{(B)}_{j'} s_{\nu,\tau,j'}) \right) \right. \right. \nonumber \\
&\left. \left. + \sum_{\tau=1}^{t} J_j s^{(A)}_{\nu,\tau,j}s^{(B)}_{\nu,\tau,j+1} + \sum_{\tau=1}^{t-1} \frac{\pi}{4} (s^{(A)}_{\nu,\tau,j}s^{(A)}_{\nu,\tau+1,j}+s^{(B)}_{\nu,\tau,j'}s^{(B)}_{\nu,\tau+1,j'}) \right) \right], \nonumber \\ 
& \times \prod_{\nu=1}^{n} \left( \frac{1+s^{(A)}_{\nu,t,j}s^{(A)}_{n+1+\mathrm{mod}(\nu-2,n),t,j}}{2} \right)\left( \frac{1+s^{(B)}_{\nu,t,j}s^{(B)}_{n+1+\mathrm{mod}(\nu-2,n),t,j}}{2} \right) \nonumber\\
&\times \prod_{\nu=1}^{2n} \left[ \cos(\theta^{(A)}_{j}/2)\delta_{s_{\nu,1,j},1} + \sin(\theta^{(A)}_{j}/2)e^{i\phi^{(A)}_{j}\mathrm{sgn}(n-\nu)}\delta_{s_{\nu,1,j},-1} \right]\left[ \cos(\theta^{(B)}_{j'}/2)\delta_{s_{\nu,1,j'},1} + \sin(\theta^{(B)}_{j'}/2)e^{i\phi^{(B)}_{j'}\mathrm{sgn}(n-\nu)}\delta_{s_{\nu,1,j'},-1} \right]. \label{eq:rmatelement}
\end{align}

Using the above matrix elements, we thus obtain 
\begin{equation} \label{seq:trrhotr}
\tr[(\rho_{X}(t))^{n}]= \tr\left[ \left( \prod_{j=1}^{N} \mathbb{T}[\theta_{j},\phi_{j}]\right) \left(\prod_{j=N+1}^{L} \mathbb{R}[\theta_{j},\phi_{j}]\right) \right].
\end{equation}

Further writing the Eq. \eqref{eq:tmatelement} in matrix form we have
\begin{align}
    \mathbb{T}[\theta,\phi]&= \prod_{\nu=1}^{n} \mathbb{T}^{(\nu)}[\theta,\phi] \\
    \mathbb{R}[\theta,\phi] &= \mathbb{P}\mathbb{T}[\theta,\phi]\mathbb{P}^{\dagger}
\end{align}
and $\mathbb{P}$ is 
\begin{align}
    \mathbb{P}= \left(\mathbb{1} \otimes \prod_{\nu=1}^{n}\prod_{\tau=1}^{t} P^{(A)}_{(\nu,\tau),(\nu-1,\tau)} \right) \otimes \left(\mathbb{1} \otimes \prod_{\nu=1}^{n}\prod_{\tau=1}^{t} P^{(B)}_{(\nu,\tau),(\nu-1,\tau)}\right),
\end{align}
and $P_{(\nu,\tau),(\nu-1,\tau)}$ is an elementary transposition
\begin{equation}
   P^{(\mathcal{A})}_{(\nu,\tau),(\nu-1,\tau)}= \frac{1}{2}\mathbb{1}+ \frac{1}{2}\sum_{\beta\in\{x,y,z\}}\sigma^{(\beta,\mathcal{A})}_{\nu,\tau}\sigma^{(\beta,\mathcal{A})}_{\nu',\tau'}.
\end{equation}
The matrix $\mathbb{T}^{(\nu)}[\theta,\phi]$  acts non-trivially only on the $\nu$-th copy and is explicitly given as product of a Hermitian, projection and a unitary operator.
\begin{align}
    \mathbb{T}^{(\nu)}[\theta,\phi]= \mathbb{H}^{z}_{\nu,1}.\mathbb{G}^{z}_{\nu,t}.\mathbb{U}^{(\nu)}[h],
\end{align}
where 

\begin{align}\label{suppeq:hguform}
  \mathbb{H}^{a}_{\nu,1}&= (2 [\cos(\theta^{(A)}/2) P^{a,+,A}_{\nu,\tau}+\sin(\theta^{(A)}/2)P^{a,-,A}_{\nu,\tau}]^{\otimes 2}) \otimes(2 [\cos(\theta^{(B)}/2) P^{a,+,B}_{\nu,\tau}+\sin(\theta^{(B)}/2)P^{a,-,B}_{\nu,\tau}]^{\otimes 2} ), \\
   \mathbb{G}^{a}_{\nu,t}&= \left(\frac{1}{2}(\mathbb{1}+ \sigma_{\nu,\tau}^{a,A}\otimes \sigma_{\nu,\tau}^{a,A})\right)\otimes \left(\frac{1}{2}(\mathbb{1}+ \sigma_{\nu,\tau}^{a,B}\otimes \sigma_{\nu,\tau}^{a,B})\right), \\
   \mathbb{U}{(\nu)}[h] &\equiv (U_{AB}U_{\nu,\phi}e^{-i h M_{\nu}^{z}} e^{i \frac{\pi}{4}  M_{\nu}^{x}} \otimes U^{*}_{AB}U^{*}_{\nu,\phi}e^{i h M_{\nu}^{z}}e^{-i \frac{\pi}{4} M_{\nu}^{x}}),
\end{align}
where
\begin{align}
   U_{\nu,\phi} &\equiv \exp\left[-\frac{i\pi}{4}\sum_{\tau=1}^{t-1}\sigma^{z,A}_{\nu,\tau}\sigma^{z,A}_{\nu,\tau+1} -i\frac{\phi}{2}\sigma^{z,A}_{\nu,1}  \right] \otimes  \exp\left[-\frac{i\pi}{4}\sum_{\tau=1}^{t-1}\sigma^{z,B}_{\nu,\tau}\sigma^{z,B}_{\nu,\tau+1} -i\frac{\phi}{2}\sigma^{z,B}_{\nu,1}  \right], \\
   U_{AB} &= \exp\left[ -i \sum_{\tau=1}^{t}J_{j}\sigma^{z,A}_{\nu,\tau,j}\sigma^{z,B}_{\nu,\tau,j+1}\right],\\
   M_{\nu}^{a} &\equiv \left(\sum_{\tau=1}^{t}\sigma_{\nu,\tau}^{a,A}\right) \otimes \left(\sum_{\tau=1}^{t}\sigma_{\nu,\tau}^{a,B}\right),\\
   P^{a,\pm,A/B}_{\nu,\tau} &\equiv \frac{1}{2}(\mathbb{1}\pm \sigma_{\nu,\tau}^{a,A/B}).
\end{align}

Notice that the structure of all the matrices involved is $A^{+}\otimes A^{-} \otimes B^{+}\otimes B^{-}$ while as written, $U_{AB}$ inherits the structure $A^{+} \otimes B^{+}\otimes A^{-} \otimes B^{-}$. Here $\pm$ denotes the positive and negative times copy respectively. This remark is specifically important while doing the numerical calculations. 


\section{Proof of Property 2}\label{suppsec:propproof}

In this section we now prove the properties of the transfer matrix

\begin{prop*}
    The following properties hold 
    \begin{enumerate}
        \item $|\lambda_{i}| \leq \lambda_{max}^{(AB)} \equiv (1+|\cos(\theta)|)^{2n}= \lambda_{max}^{(A)}\lambda_{max}^{(B)}\,\,\forall{} \lambda_{i}\in\textrm{Spec}(\mathbb{T}).$
        \item For an eigenvalue $\lambda$ of $\mathbb{T}$ that satisfies $|\lambda|= \lambda_{\max}$ we have
        \begin{enumerate}
            \item[(a)] The geometric and algebraic multiplicities of $\lambda$ coincide i.e. it has trivial Jordan blocks. 
            \item[(b)] the left eigenvector $\bra{A}$ then satisfies
            $$\bra{A}\prod_{\nu=1}^{n}\mathbb{H}_{\nu,1}^{z} = \lambda_{max} \bra{A},$$
            $$\bra{A}\prod_{\nu=1}^{n}\mathbb{G}^{z}_{\nu,1}= \bra{A},$$
            $$\bra{A}\prod_{\nu=1}^{n}\mathbb{U}= e^{i\alpha}\bra{A}.$$
        \end{enumerate}
 \end{enumerate}
\end{prop*}

\emph{Proof.--} For a generic state $\bra{A}$ and using the fact that $\mathbb{G}_{\nu,t}^{z}$ is a projector and its expectation value on a normalized state is less than or equal to 1, we can write
\begin{equation}
    \braket{A|\mathbb{T}\mathbb{T}^{\dagger}|A}= \braket{A|\prod_{\nu=1}^{n}\mathbb{H}_{\nu,1}^{z} \prod_{\nu=1}^{n}\mathbb{G}_{\nu,t}^{z} \prod_{\nu=1}^{n} \mathbb{H}_{\nu,1}^{z}|A} \leq \braket{A|\prod_{\nu=1}^{n}(\mathbb{H}_{\nu,1}^{z})^{2} |A}.
\end{equation}
Now, using Eq. \eqref{suppeq:hguform} we notice that $\mathbb{H}_{\nu,1}^{z}$ has the form $\mathbb{H}_{\nu,1}^{z,A} \otimes \mathbb{H}_{\nu,1}^{z,B}$ which implies that the result for the full chain factorizes into result of two dual chains. Hence we can use the result of single dual-unitary chain \cite{Bertini:2018fbz} and obtain
\begin{equation}
    \lambda_{max}= \lambda_{max}^{(A)}\lambda_{max}^{(B)}= (1+ |\cos(\theta)|)^{n} (1+ |\cos(\theta)|)^{n}=(1+ |\cos(\theta)|)^{2n},
\end{equation}
This proves property \emph{1}. 
In fact looking at the explicit form of operators as given by Eq. \eqref{suppeq:hguform}, we can obtain all the other properties by merely using the properties of single dual chain \cite{Bertini:2018fbz}. Note that although $ \mathbb{U}{(\nu)}[h]$ has a part, $U_{AB}$, that is non-factorizable as the contribution from two sublattices, it only affects the phase $e^{i \alpha}$.
\section{Numerics using duality mapping}\label{suppsec:duality}
In this section we briefly discuss the details of numerics done using duality mapping. For this we mainly follow the method outline in \cite{Bertini:2018fbz}. First we note that using Eq. \eqref{seq:trrhotr} we can write 
\begin{align}
    S^{(n)}_{X}(t)= \frac{1}{1-n} \ln \left[\tr\left[ \left( \prod_{j=1}^{N} \mathbb{T}[\theta_{j},\phi_{j}]\right)\mathbb{P} \left(\prod_{j=N+1}^{L} \mathbb{T}[\theta_{j},\phi_{j}]\right) \mathbb{P}^{\dagger}\right]\right].
\end{align}

For simplicity we consider
$$h_{j}= h, \quad \theta_{j}=\theta,\quad \phi_{j}=\phi.$$

Then taking the thermodynamic limit, we have 
\begin{equation}\label{seq:dualrel1}
    \lim_{N \rightarrow \infty }\lim_{L \rightarrow \infty } S^{(n)}_{X}(t)= \frac{2}{1-n} \log \left| \frac{\braket{V_{L}|\mathbb{P}|\mathbb{V_{R}}}}{\braket{V_{L}|\mathbb{V_{R}}}} \right|,
\end{equation}
where $\bra{V_{L}}$ and $\ket{V_{L}}$ are left and right eigenstate respectively, of $\mathbb{T}[\theta,\phi]$, corresponding to eigenvalue 1. Furthermore, due to tensor product structure of the transfer matrix, we only need to search for the eigenvector of the form
\begin{equation}
    \ket{V_{R}}= \bigotimes_{\nu=1}^{n} \ket{R} \quad \bra{V_{L}}= \bigotimes_{\nu=1}^{n} \bra{L},
\end{equation}

This combined with Eq. \eqref{seq:dualrel1} gives
\begin{equation}
    \lim_{N \rightarrow }\lim_{L \rightarrow } S^{(n)}_{X}(t)= \frac{2}{1-n} \log  \frac{\tr[(R^{\dagger}L)^{n}]}{[\tr(R^{\dagger}L)]^{n}}.
\end{equation}
We can easily generalize the method to the case when we have two \emph{uniform} species such as 
$$h_{j}= h, \quad \theta^{(A)}_{j}=\theta,\quad \phi^{(A)}_{j}=\phi, \quad \theta^{(B)}_{j}=\theta,\quad \phi^{(B)}_{j}=\phi. $$

\section{A minimal model with heterogeneous local Hilbert space}\label{suppsec:d2d3}
We now give a concrete realization of a model with heterogeneous local Hilbert spaces. For this we will consider two types of dual models on sublattice $A$ and $B$.
For lattice $A$ we will consider the KFIM model, with local dimension $d=2$, for which we have 
\begin{align}
    H_{I}^{AA} &= J_{A} \sum_{n\in A}\sigma^{z}_{n}\sigma^{z}_{n+2} +\sum_{n \in A}h^{(A)}_{n}\sigma_{n}^{z},\\
    H^A_{K} &= b_{A} \sum_{n \in A}\sigma^{x}_{n}.\\
\end{align}

For sublattice $B$ we take the Potts model \cite{PhysRevB.105.144306,Claeys:2024tuy}, with local dimension $d=3$. For this we consider the following Hamiltonian:
\begin{align}\label{seqn:kpm}
    H^{BB}_{I}& =  J_{B} \sum_{n \in B} ({Z}_{n}{Z}^{\dagger}_{n+1} +{Z}^{\dagger}_{n}{Z}_{n+1})+ \sum_{n \in B}h^{(B)}_{n}(Z_{n}+ Z^{\dagger}_{n})\nonumber \\
    H^{B}_{K}& = b_{B}\sum_{n \in B} \, (X_{n}+X^{\dagger}_{n}).
\end{align}
Denoting $\omega= e^{2 \pi i/3}$, the explicit representation of $Z$ and $X$ is 
\begin{equation}\label{seqn:zqxq}
Z=\left(\begin{array}{cccc}
1 & 0 & 0 \\
0 & \omega & 0  \\
0 & 0 &\omega^2
\end{array}\right), \quad X=\left(\begin{array}{ccccc}
0 & 1 & 0  \\
0 & 0 & 1\\
1 & 0 & 0 
\end{array}\right) .
\end{equation}
with $X^{3} = Z^{3}= I$.
We can then have the inter-lattice couplings as 
\begin{equation}
    H^{AB}= \sum_{n=1}^{L} J_{n}(\sigma_{n}Z_{n+1}+\sigma_{n}Z^{\dagger}_{n+1})
\end{equation}
This model has dual unitarity for $|J_{A}|= |b_{A}|= \frac{\pi}{4}$ and $|J_{B}|= |b_{B}|= \frac{2\pi}{3},\frac{4\pi}{3}$. All the other coupling $J_{n},\, h_{n}^{(A)}$ and $h_{n}^{(B)}$ can be generic.
\bibliography{references}